\documentclass[11pt]{article}

\input epsf
\usepackage{amsmath,epsfig,placeins,subfigure}
\usepackage[perpage,symbol*]{footmisc}

\title{\bf Properties of four numerical schemes applied to a nonlinear scalar wave equation  
with a GR-type nonlinearity}  

\author{ Jakob Hansen\footnotemark[1],
Alexei Khokhlov\footnotemark[2] \ and Igor Novikov\footnotemark[1] 
{\footnotesize $^,$}\footnotemark[3]
{\footnotesize $^,$}\footnotemark[4] }

\begin{document}

\footnotetext[1]{Niels Bohr Institute, Blegdamsvej 17, DK-2100 Copenhagen, Denmark}
\footnotetext[2]{\mbox{Department of Astronomy and Astrophysics, The University of Chicago, 5640 Ellis
    Avenue, Chicago, IL 60637, USA}} 
\footnotetext[3]{NORDITA, Blegdamsvej 17, DK-2100  Copenhagen, Denmark}
\footnotetext[4]{Astro Space Center of P.N. Lebedev Physical Institute, Profsoyouznaja 83/32, Moscow
     118710, Russia}

\maketitle

\large

\begin{abstract}
We study stability, dispersion and dissipation properties of four numerical schemes (Iterative
Crank-Nicolson, 3'rd and 4'th order Runge-Kutta and Courant-Fredrichs-Levy Non-linear). By use of a
Von Neumann analysis we study the
schemes applied to a scalar linear wave equation as well as a scalar non-linear wave equation with a
type of non-linearity present in GR-equations. Numerical testing is
done to verify analytic results. We find that the method of lines (MOL) schemes 
are the most dispersive and dissipative schemes. 
The
Courant-Fredrichs-Levy Non-linear (CFLN) scheme is most accurate and 
least dispersive and dissipative, but the absence of 
dissipation at Nyquist frequency, if fact,  puts it at a disadvantage in numerical simulation.
Overall, the  4'th
order Runge-Kutta scheme, which has the least amount of dissipation 
among the MOL schemes, seems to be the most suitable compromise between the overall accuracy and 
damping at short wavelengths.
\end{abstract}

\clearpage
\numberwithin{equation}{section}

\section{\bf  Introduction}
\label{sec:1}

The area of numerical relativity is very complex with many factors influencing the
calculations. Due to this complexity, choosing a best numerical scheme for solving
problems in general relativity (GR) may be of particular
importance in order to have a long term stability and to minimize truncation errors. At the 
same time, problems  in numerical relativity usually demand a large amount of computational
resources. Hence it is 
also desirable to choose a scheme which best exploits the available computer power at any given
time, i.e. provides accurate results with a minimal number of operations.

First and foremost the scheme needs to be numerically stable,
but in addition to this basic (but not always trivially established) property, there are a number of
other features, such as accuracy, dissipation and dispersion properties, which may have important
impacts on the numerical solutions as well.

The purpose of this paper is to investigate numerical properties (in particular dispersion, stability and
dissipation) of some numerical schemes which can be used to solve GR type equations, allowing users
of these schemes to estimate potential pitfalls and limitations of the schemes. The schemes that
are being analyzed are the 
iterative Crank-Nicolson scheme (ICN), the third and fourth order Runge-Kutta schemes (RK3 and RK4)
and a nonlinear version of the classical Courant-Friedrichs-Levy
scheme (CFLN). 
A unique aspect of numerical GR is that the
integration may be affected not only by stability of the scheme, but also by constraint and gauge
instabilities which are intrinsic to the equations themselves.
In order to separate these effects
from the stability of the schemes and to investigate strengths and weaknesses of the different
schemes, we want to apply the schemes to simpler scalar wave equations.
Numerical schemes are usually analyzed by their application to the
standard linear scalar wave equation (e.g. \cite{Alcubierre}, \cite{Teukolsky})
however, they are usually applied to solve nonlinear problems. By use of a classical Von Neumann
analysis, we analyze properties of the schemes applied to both the standard scalar linear wave equation and 
to a nonlinear scalar wave equation, which is designed to mimic some
properties of the nonlinear terms in the Einstein equations. By this we hope to expand our knowledge
of the behavior of the schemes in a regime which is closer to the computational reality of numerical
GR then a sole analysis of the schemes in the fully linear regime.
The non-linear wave equation, a perturbation analysis of it and the numerical schemes are presented
in section \ref{sec:theory}.
The results from the Von Neumann analysis of the schemes applied to the scalar linear and nonlinear
wave equations are presented 
in section \ref{sec:linear} and \ref{sec:nlwave} respectively. To support the analytic results,
numerical results are presented in section \ref{sec:4}.

\section{\bf  Formulation of the problem}
\label{sec:theory}
In this section we present the numerical schemes which we will analyze in subsequent sections, the
non-linear scalar wave equation to which we will apply the schemes and the method by which we
will be analysis the schemes. 
\subsection{\bf  A Non-linear Wave Equation}
\label{subsec:nlwave}
Consider the following scalar, quasi-linear, hyperbolic partial differential equation of two
independent variables, $t$ and $x$ (\cite{Khoklov}):
\begin{equation}\label{eq:nlwe}
  \frac{\partial^2 g}{\partial t^2} =  \frac{\partial^2 g}{\partial x^2}-\frac{1}{g}\left(
  \frac{\partial g}{\partial t}\right)^2
\end{equation}
or cast into first order form:
\begin{equation}\label{eq:nlwe2}
\begin{split}
  \frac{\partial g}{\partial t} &=  K\\
  \frac{\partial K}{\partial t} &=\frac{\partial^2 g}{\partial x^2}-\frac{\left( K\right)^2}{g}
\end{split}
\end{equation}
This equation is essentially the standard scalar wave equation with an added non-linear term. The
equation is interesting because the non-linear term mimics part of the non-linearities present in GR
equations. This can be seen by recalling that the structure of a Ricci tensor $R_{ab}$ is $R\sim
\sum \partial \Gamma + \sum\Gamma\Gamma$, where Christoffel symbols $\Gamma \sim g^{-1}\partial g$,
and $g$ is the metric. Thus $R_{ab}$ can be represented as a sum of the terms $R\sim \sum
g^{-1}\partial^2 g + \sum g^{-2}(\partial g)^2$. Equation \eqref{eq:nlwe} thus mimics a type of
non-linearity present in GR equations $R_{ab}=0$. Especially equations \eqref{eq:nlwe2} resembles
the evolutionary part of GR equation in a standard ADM 3+1 form (with zero shift and constant lapse).
We expect 
that in order to successfully apply the schemes to GR equations they must be applicable
equation \eqref{eq:nlwe}. 
 
Equation \eqref{eq:nlwe} posses a large number of non-trivial solutions \cite{Khoklov}. For test
purposes we use two particular analytical solutions to this equation:
A spatial constant solution: 
\begin{equation}\label{eq:constant}
g(t) =  \sqrt{C_1+C_2\cdot t}, (C_1+C_2\cdot t \ge 0).\footnote{$g(t)=-\sqrt{C_1+C_2\cdot t}$ is naturally
  also a solution but since we are trying to mimic some features of the Einstein equations, $g$
  (which mimics the 3-metric of the general relativity) cannot be negative. This is also why we
  require $C_1+C_2\cdot t \ge 0$ as we would otherwise obtain complex solutions which would be
  unphysical.} 
\end{equation}
and an exponential solution:
\begin{equation}\label{eq:exponential}
g(x,t)=\exp{\left(\pm\sqrt{C}\cdot x \pm \sqrt{\frac{C}{2}}\cdot
    t\right)}, (C\ge 0, \pm \mbox{ signs independent})\footnote{$C$ cannot be negative as this would
    create complex solutions 
    (see previous footnote).}.
\end{equation}

\subsection{\bf  Perturbation analysis}
\label{sec:linear_analysis}
We wish to analyze the analytical behavior of small amplitude perturbations of solutions
\eqref{eq:constant} and \eqref{eq:exponential}.

Let $g_0(x,t)$ be a base solution of equation \eqref{eq:nlwe} and let
$g_0 +\tilde{g}$ be a perturbed solution, with $|\tilde{g}|\ll |g_0|$. We linearize
\eqref{eq:nlwe} around $g_0$ and obtain a linear equation for the perturbations
\begin{equation}
  \label{eq:linearized_nlwe}
  \frac{\partial^2 \tilde{g}}{\partial t^2}=\frac{\partial^2 \tilde{g}}{\partial x^2}-2\left(
  \frac{1}{g_0}\frac{\partial g_0}{\partial t}\right)\frac{\partial \tilde{g}}{\partial t}+\left(
  \frac{1}{g_0}\frac{\partial g_0}{\partial t}\right)^2 \tilde{g}
\end{equation}
or
\begin{equation}
  \frac{\partial^2 \tilde{g}}{\partial t^2}=\frac{\partial^2 \tilde{g}}{\partial
  x^2}-2A\frac{\partial \tilde{g}}{\partial t}+A^2 \tilde{g}   
\end{equation}
where
\begin{equation}
  A\equiv \frac{1}{g_0}\frac{\partial g_0}{\partial t}
\end{equation}
Consider perturbations of the form:
\begin{equation}
  \tilde{g}\propto e^{I\omega t-Ikx}.
\end{equation}
where $I=\sqrt{-1}$. Then the dispersion relation is
\begin{equation}\label{eq:disp}
  \omega^2 = k^2 + 2AI\omega - A^2.
\end{equation}
Substituting $\omega =\omega_R + I\omega_I$ into equation \eqref{eq:disp} and separating real and
imaginary parts we obtain:
\begin{equation}\label{eq:imag}
  2I\omega_R\omega_I = 2IA\omega_R
\end{equation}
\begin{equation}\label{eq:real}
  \omega_R^2 -\omega_I^2 = k^2-2A\omega_I - A^2
\end{equation}
From \eqref{eq:imag} it follows that $\omega_I = A$,
hence the growth rate of a perturbation $\tilde{g}\propto e^{I\cdot I\omega_I t}$ is 
\begin{equation}\label{eq:perturbation_analysis_amp}
  \tilde{g}\propto e^{-A t}.
\end{equation}
We see that a base solution is stable if $A \ge 0$ and otherwise it is unstable. 
Hence growing solutions are
stable and decaying solutions are unstable.

Due to a local
linearizion involved in the above analysis, the conclusion is strictly valid for 
perturbations with high 
wave numbers $k=\frac{2\pi}{\lambda}$ compared to the base solutions, and only if the growth rate is
$\gg$ then that of the base solution. By looking at the spatial constant base solution
\eqref{eq:constant}, perturbed by a spatial constant perturbation ($k=0$), we can
estimate the behavior of spatial constant or very long 
wavelength perturbations. 

Consider a spatial constant base solution \eqref{eq:constant} perturbed by a small spatially constant
perturbation: $g(t) = g_0(t) + \tilde{g}(t)$. Then equation
\eqref{eq:linearized_nlwe} simplifies to an 
ordinary differential equation which can be solved to yield:
\begin{equation}\label{eq:constantperturbationsol}
  \tilde{g}(t) = \pm \frac{K}{2}\left(\frac{\sqrt{2}}{g_0(t)}+\frac{g_0(t)}{\sqrt{2}} \right)
\end{equation}
where $K$ is an arbitrary integration constant. From \eqref{eq:constantperturbationsol}
 we see that for growing base solutions
(i.e positive $C_2$ in \eqref{eq:constant}) the second term is growing proportionally to the base
solution, while the 
first term is decaying inversely proportionally to the base solution, hence the relative error is
decreasing towards a constant value proportional to $K$. That is, the perturbation is growing but 
it is not unstable. 
If the base solution is decaying (i.e. negative $C_2$ in \eqref{eq:constant}), the first 
term increases rapidly as $g_0(t)\rightarrow 0$ and the solution becomes unstable as the relative
error grows unbounded.

\subsection{\bf  The Numerical Schemes}
\label{subsec:schemes}
The four schemes that we are investigating are 
\begin{itemize}
\item Iterative Crank-Nicolson
\item 3'rd order Runge-Kutta
\item 4'th order Runge-Kutta
\item Courant-Friedrichs-Levy Nonlinear 
\end{itemize}
The first three schemes are based on the method of lines
approach \cite{MOL}, while the Courant-Friedrichs-Levy Nonlinear scheme is based on the classic
central-difference, second-order explicit scheme introduced in 1928 by Courant,
Friedrichs and Levy \cite{Courant}. 
The schemes are defined as follows:

\subsubsection{\bf  The Iterative Crank-Nicolson scheme} 
The iterative Crank-Nicolson scheme (ICN) is an explicit, iterative scheme which was developed by
Matt Choptuik from the classic 
implicit Crank-Nicolson scheme \cite{Crank}, \cite{Richtmeyer}. To solve equtation \eqref{eq:nlwe2},
we define the
ICN-scheme as follows. 
First the iteration process is initiated:
\begin{equation}
\begin{split}
  K_i^{(1)} &= K_i^n + \Delta t \left(\delta^2 \left( g_{i}^n\right) + NLT\left(
  g_i^n,K_i^n\right)\right)\\  
  g_i^{(1)} &= g_i^n + \Delta t \cdot K_i^n 
\end{split}  
\end{equation}
where $\delta^2 (g_i^n)=\frac{g_{i-1}^n-2g_i^n+g_{i+1}^2}{\Delta x^2}$ is the centered second order
accurate finite difference approximation to the second order spatial derivative,  $g_i^n$ and
$K_i^n$ are determined at mesh points $x_i =i\Delta x$, $t^n = n\Delta t$ and $NLT\left(
  g_i^n,K_i^n\right)$ is the non-linear term (i.e for eq. \eqref{eq:nlwe2} $NLT\left(
  g_i^n,K_i^n\right) = \frac{(K^i_n)^2}{g^i_n}$). The scheme is then
iterated: 
\begin{equation}
\begin{split}
  K_i^{(j)} &= K_i^n + \Delta t \left(\delta^2 \left( \frac{g_i^{(j-1)}+g_i^n}{2} \right)
  +NLT\left(\frac{g_i^{(j-1)}+g_i^n}{2},\frac{K_i^{(j-1)}+K_i^n}{2}\right)\right)\\ 
  g_i^{(j)} &= g_i^n + \Delta t\left( \frac{ K_i^{(j-1)} + K_i^n}{2} \right)
\end{split}  
\end{equation}
$j\in [2,j_{max}]$), and  finally the dependent variables at the next time step are:
\begin{equation}
  \begin{split}
      K_i^{(n+1)} &= K_i^{(j_{max})}\\
      g_i^{(n+1)} &= g_i^{(j_{max})}
  \end{split}
\end{equation}
As shown in \cite{Alcubierre} and \cite{Teukolsky}, the optimal number of iterations, for the scalar
wave equation is $j_{max}=3$\footnote{Note that different authors count the number of iterations in
  different ways, some do not count the first step as an iteration and hence state that the optimal
  number of iterations is 2.}, which is the scheme that we
will investigate in this paper. This 
scheme can be shown to be second order accurate in both time and space by a Taylor series
expansion \cite{Alcubierre},\cite{Teukolsky}.

\subsubsection{\bf  The 3'rd order Runge-Kutta scheme}
To solve eq. \eqref{eq:nlwe2}, the 3'rd order Runge-Kutta scheme (RK3) is defined as
follows \cite{Abramowitz}: 
\begin{equation}
  \begin{split}
    K_i^{n+1} &= K_i^n + \frac{K^{(1)}_i+4K^{(2)}_i+K^{(3)}_i}{6}\\
    g_i^{n+1} &= g_i^n + \frac{g^{(1)}_i+4g^{(2)}_i+g^{(3)}_i}{6}  
  \end{split}
\end{equation}
where 
\begin{equation}
  \begin{split}
    K^{(1)}_i &= \Delta t \left[ \delta^2 \left( g_i^n \right) +NLT\left( g_i^n, K_i^n\right)\right]\\
    g^{(1)}_i &= \Delta t \left[  K_i^n\right]\\
    K^{(2)}_i &= \Delta t \left[ \delta^2 \left( g_i^n + \frac{g^{(1)}_i}{2}\right) +NLT\left( g_i^n+
    \frac{g^{(1)}_i}{2}, K_i^n+ \frac{K^{(1)}_i}{2} \right)\right]\\ 
    g^{(2)}_i &= \Delta t \left[  K_i^n+ \frac{K^{(1)}_i}{2}\right]\\
    K^{(3)}_i &= \Delta t \left[ \delta^2 \left( g_i^n - g^{(1)}_i + 2g^{(2)}_i\right) +NLT\left(
    g_i^n- 
    g^{(1)}_i + 2g^{(2)}_i, K_i^n- K^{(1)}_i + 2K^{(2)}_i \right)\right]\\
    g^{(3)}_i &= \Delta t \left[  K_i^n- K^{(1)}_i + 2K^{(2)}_i \right]
  \end{split}
\end{equation}
This scheme can be shown to be second order accurate in space and third order accurate in time by a
Taylor series expansion.

\subsubsection{\bf  The 4'th order Runge-Kutta scheme}
To solve eq. \eqref{eq:nlwe2} the 4'th order Runge-Kutta scheme (RK4) is defined as
follows \cite{Abramowitz}: 
\begin{equation}
  \begin{split}
    K_i^{n+1} &= K_i^n + \frac{K^{(1)}_i+2K^{(2)}_i+2K^{(3)}_i+K^{(4)}_i}{6}\\
    g_i^{n+1} &= g_i^n + \frac{g^{(1)}_i+2g^{(2)}_i+2g^{(3)}_i+g^{(4)}_i}{6}  
  \end{split}
\end{equation}
where 
\begin{equation}
  \begin{split}
    K^{(1)}_i &= \Delta t \left[ \delta^2 \left( g_i^n \right) +NLT\left( g_i^n, K_i^n\right)\right]\\
    g^{(1)}_i &= \Delta t \left[  K_i^n\right]\\
    K^{(2)}_i &= \Delta t \left[ \delta^2 \left( g_i^n + \frac{g^{(1)}_i}{2}\right) +NLT\left( g_i^n+
    \frac{g^{(1)}_i}{2}, K_i^n+ \frac{K^{(1)}_i}{2} \right)\right]\\ 
    g^{(2)}_i &= \Delta t \left[  K_i^n+ \frac{K^{(1)}_i}{2}\right]\\
    K^{(3)}_i &= \Delta t \left[ \delta^2 \left( g_i^n + \frac{g^{(2)}_i}{2}\right) +NLT\left( g_i^n+
    \frac{g^{(2)}_i}{2}, K_i^n+ \frac{K^{(2)}_i}{2} \right)\right]\\ 
    g^{(3)}_i &= \Delta t \left[  K_i^n+ \frac{K^{(2)}_i}{2}\right]\\
    K^{(4)}_i &= \Delta t \left[ \delta^2 \left( g_i^n + g^{(3)}_i\right) +NLT\left(
    g_i^n+g^{(3)}_i, K_i^n + K^{(3)}_i \right)\right]\\
    g^{(4)}_i &= \Delta t \left[  K_i^n +K^{(3)}_i \right]
  \end{split}
\end{equation}
This scheme can be shown to be second order accurate in space and fourth order accurate in time by a
Taylor series expansion.

\subsubsection{\bf  Crank-Friedrichs-Levy Nonlinear scheme (CFLN)}
Another approach to solving eq.\eqref{eq:nlwe2} numerically is to notice that without the non-linear
term, equation \eqref{eq:nlwe2} is a
scalar wave equation which can be solved by the classic explicit scheme by Courant,
Friedrichs and Levy \cite{Courant}(see also cp. 10 in \cite{Richtmeyer}):
\begin{equation}
  \frac{g_i^{n+1}-2g_i^{n}+g_i^{n-1}}{\Delta t^2}=\frac{g_{i+1}^{n}-2g_{i}^{n}+g_{i-1}^{n}}{\Delta x^2}
\end{equation}
We can cast this scheme into a first order form:
\begin{equation}\label{eq:3}
\begin{split}
 K_i^{n+\frac{1}{2}} &= K_i^{n-\frac{1}{2}}+\Delta
 t\left(\delta^2 (g_i^n)\right) \\ 
g_i^{n+1} &= g_i^{n}+\Delta t K_i^{n+\frac{1}{2}}
\end{split}
\end{equation}
Now, in order to use this scheme as a basis for solving equation \eqref{eq:nlwe2}, we must add a term to
evolve the non-linear part, moreover, we wish to do this with second-order accuracy at the grid
points ($x_i,t^n$) to ensure overall second-order accuracy of the scheme. We do this by evaluating
the non-linear term using the following predictor-corrector style approach \cite{Khoklov}:
\begin{equation}\label{eq:4}
\begin{split}
\tilde{K}_i^{n+\frac{1}{2}}&=K_i^{n-\frac{1}{2}}+\Delta
t\left(\delta^2 (g_i^n) +NLT(g_i^n,K_i^{n-\frac{1}{2}})\right)\\
K_i^{n+\frac{1}{2}}&=\tilde{K}_i^{n+\frac{1}{2}}+\frac{\Delta t}{2}\left( NLT(g_i^n,
  K_i^{n-\frac{1}{2}})+ NLT(g_i^n, \tilde{K}_i^{n+\frac{1}{2}}\right)\\ 
g_i^{n+1} &= g_i^n + \Delta t K_i^{n+\frac{1}{2}}
\end{split}
\end{equation}

This scheme can be shown to be second order accurate in both space and time by a Taylor series expansion. An 
advantage of the scheme is that the second-order 
accuracy is achieved with a relatively small number of right hand side operations, to calculate one
time step CFLN requires 2 evaluations of the non-linear terms, compared to 3 evaluations required for
the ICN and RK3 schemes and 4 evaluations required for the RK4 scheme. It is also noted that the
scheme is staggered in time, i.e. initial values are required at points
$g^n_{j-i,j,j+1},K^{n-\frac{1}{2}}_i$. 

\subsection{\bf  The Von Neumann analysis}
To investigate the properties of the numerical schemes we use a Von Neumann analysis
 following a standard approach \cite{Richtmeyer},\cite{Miller}:
All the schemes can be represented by the following evolution operator:
\begin{equation}\label{eq:pavel1}
  {U}^{n+1}_{j,l} = S^n_{j,l} ({U}^n_{j',l'})
\end{equation}
where ${U}^n_{j,l}$ is a set of dynamical variables, $n$ and $j$ are temporal and spatial indices
 respectively and $l$ enumerates the dynamic variables. In our case, the
operator $S^n_{j,l}$ may depend on all components of
${U}^n_{j',l'}$ at grid
points ($j=j-1,j,j+1$) corresponding to a time layer $n$ (or a combination of $n$ and
 $n-\frac{1}{2}$ in the staggered case of CFLN scheme). 

Perturbing ${U}^n_{j,l}$ as
\begin{equation}\label{eq:pavel2}
  {U}^n_{j,l}=\hat{U}^n_{j,l} + \delta {U}^n_{j,l},
\end{equation}
where $\hat{U}^n_{j,l}$ is the background solution, substituting equation
\eqref{eq:pavel2} into \eqref{eq:pavel1} and doing a Taylor-expansion, we obtain

\begin{equation}\label{eq:pavel3.2}
  \delta {U}^{n+1}_{j,l} = \sum_{j',l'} \frac{\partial S^n_{j,l}
   }{\partial \hat{U}^n_{j',l'}} \delta {U}^n_{j',l'}+ O\left( \left( {\delta
        {U}^n_{j',l'}}\right)^2 \right)
\end{equation}
Assuming a perturbation of the form
\begin{equation}\label{eq:pavel4}
  \delta {U}^n_{j,l}=\xi^n_l e^{-Ij\Delta x k},
\end{equation} where $0<k<\frac{\pi}{\Delta x}$ is the perturbation wave number and $I=\sqrt{-1}$,
and substituting \eqref{eq:pavel4} into 
\eqref{eq:pavel3.2} we obtain
\begin{equation}\label{eq:pavel5}
{\xi}^{n+1}_l \approx \sum_{j',l'}\frac{\partial
  S^n_{j,l}}{\partial {U}^n_{j',l'}} \xi^{n}_{l'} e^{I(j-j')\Delta x k}\equiv G^n_ {j,l,l'}\xi^n_ {l'}
\end{equation}
The {\it amplification
  matrix\/}\index{Amplification Matrix} $G^n_{j,l,l'}$
\begin{equation}\label{eq:pavel6}
  G^n_{j,l,l'} = \sum_{j'}\frac{\partial
  S^n_{j,l}}{\partial {U}^n_{j',l'}} e^{I(j-j')\Delta x k}
\end{equation}
holds information about properties of the numerical schemes.

For a linear finite difference equation, the amplification matrix only depends on the discretization
parameters $\Delta t$, $\Delta x$ and the wave number $k$, and needs to be calculated only once in
order to know properties of the scheme at all
times and grid points. 
For non-linear schemes the amplification matrix depends on $\hat{U}^n_{j,l}$ and must in principle
be evaluated at 
\textit{all} spatial points and at \textit{all} points in time.

The generally complex eigenvalues of the amplification matrix, $\lambda_i$, hold information
about the amplification and speed of a given perturbation for a single time step  which can be
extracted by calculating the modulus and argument of the eigenvalues respectively.
The classic condition for numerical stability is that the spectral radius
(i.e. the largest modulus of the eigenvalues) of
the amplification matrix is less than or equal to $1$ for all wave numbers \cite{Richtmeyer},
however this excludes the possibility of growing exponential solutions. A less strict stability
requirement (the Von Neumann stability criterion) 
which allows for exponentially growing
solutions is that the eigenvalues must satisfy 
\begin{equation}\label{eq:stability}
  |\lambda|\equiv \max_{i}|\lambda_i|\le 1 + O(\Delta t)
\end{equation}
for all $k$ \cite{Richtmeyer}. In this paper, we
will refer to the spectral radius $|\lambda |$ as the amplification factor.
We presents amplification factors as a function of  $k\cdot\Delta x$ which runs in
the range $k\cdot\Delta x \in [0,\pi]$ with $k\cdot\Delta x  =\pi$ corresponding to the Nyquist frequency. 

It should be noted that due to the local linearizion involved in calculating the
eigenvalues for non-linear schemes, the Von Neumann analysis is only locally
valid. This also means that the analysis cannot be 
trusted for non-local wave modes, i.e. small wave numbers. However, we \textit{are} most interested in
high wave numbers, as experience tells us that numerical instabilities usually arises first at the
Nyquist frequencies. Hence for analyzing the stability of a scheme one searches for conditions under
which equation \eqref{eq:stability} is valid, focusing on the Nyquist frequency, which usually
reduces to a restriction on the relationship between $\Delta t$ and $\Delta x$, known as the
Courant number $\alpha = \frac{\Delta x}{\Delta t}$. 


\section{\bf   The linear scalar wave equation}
\label{sec:linear}

Before studying the non-linear schemes, we briefly summarize stability, dissipation and dispersion
properties of the schemes applied for a scalar linear wave equation 
\begin{equation}\label{eq:linear}
  \frac{\partial^2 g(x,t)}{\partial t^2}= \frac{\partial^2 g(x,t)}{\partial x^2}
\end{equation}

Table \ref{tab:stableinterval} shows which Courant intervals satisfies the  Von Neumann stability
criterion \eqref{eq:stability} for the Nyquist frequency as calculated by a Von Neumann analysis
in the limit $\Delta t\rightarrow 0$. 
\FloatBarrier
\begin{table}[h]\label{tab:stableinterval}
\begin{center}
\begin{tabular}{|c|c|}
\hline
 Scheme & Stable Courant interval\\
\hline
 ICN & $0<\alpha < 1$\\ %
 RK3 & $0<\alpha < \sqrt{\frac{3}{4}}$\\ 
 RK4 & $0<\alpha < \sqrt{2}$\\
 CFLN & $0 < \alpha < 1$ \\ 
 \hline
\end{tabular}
\caption{Courant intervals which satisfies the Von Neumann stability criterion \eqref{eq:stability}
  for the linear wave equation \eqref{eq:linear}}
\end{center}
\end{table}\ \\ \ \

\begin{figure}[h]
\centering
          \subfigure[Courant number $\alpha =
          0.50$]{\epsfig{figure=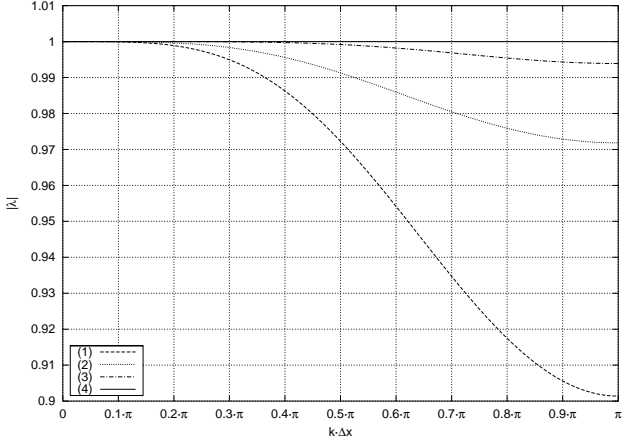,width=3.35in}\label{fig:5a}} 
          \subfigure[Courant number $\alpha =
          \sqrt{0.75}$]{\epsfig{figure=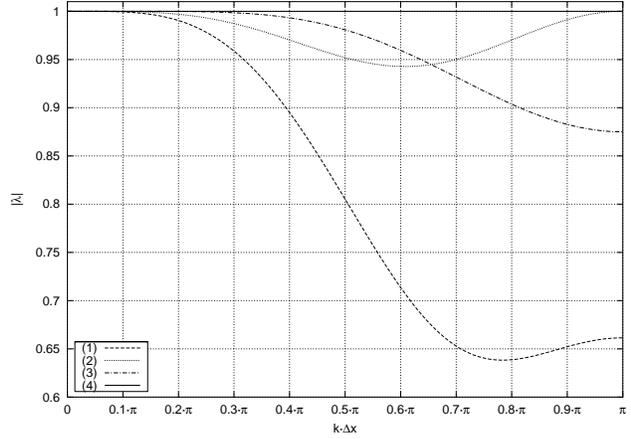,width=3.35in}\label{fig:5b}} 
\caption{{\it Amplification factor as a function of $k\cdot\Delta x$ at Courant number
          a) $\alpha =0.50$ and b) $\alpha =
          \sqrt{0.75}$, respectively, for the 4 schemes investigated. Legend is (1) = ICN, (2) = RK3, (3) = RK4 and
    (4) = CFLN. 
\label{fig:5}}}
\end{figure}

Figures \ref{fig:5} shows the amplification factors as a function of
$k\cdot\Delta x$ for Courant numbers $\alpha = 0.50$ and $\alpha = \sqrt{0.75}$ respectively, the
latter corresponds to the maximal stable Courant number for the RK3
scheme (cf. table \ref{tab:stableinterval}). These figures are representative of the
dissipative behavior of the four schemes. The plot 
(and all plots in this
section) is valid for all choices of $\Delta x$. 

\begin{figure}[h]
\begin{center}
\epsfig{figure=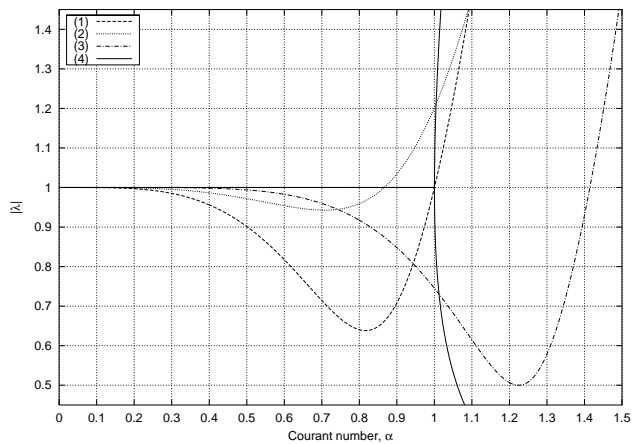,width=3.35in}
\caption[]{{\it Largest eigenvalues at the Nyquist frequency ($k\Delta x=\pi$) as function of Courant
    numbers for the 4 schemes investigated. Legend is (1) = ICN, (2) = RK3, (3) = RK4 and
    (4) = CFLN. 
\label{fig:4}}}
\end{center}
\end{figure}

For zero wave numbers all schemes are non-dissipative, but with increasing wave numbers and increasing Courant numbers
the method of lines schemes shows monotonically increasing dissipation, with ICN being the most
dissipative scheme followed by the RK3 and RK4 scheme respectively.  For very high wave numbers
the dissipation for the method of lines schemes shows a non-monotonic behavior
as can be seen by comparing figures \ref{fig:5}. We see that the ICN scheme is
still, generally, the most dissipative scheme, but as Courant numbers are increased towards the
stable limit, the dissipation at the Nyquist frequency vanishes and the maximal
dissipation is seen for smaller wave numbers. Figure \ref{fig:4} shows the dissipative behavior of
the schemes at the Nyquist frequency as a function of Courant number. This shows how the Nyquist
frequency has maximal dissipation at around $\alpha \approx 0.8\cdot
\alpha_{max}$ (where $\alpha_{max}$ is the maximal stable Courant number for the corresponding scheme), after
which the dissipation at the Nyquist frequency goes to zero just before numerical instability sets
in. 

The CFLN scheme, in contrast, is completely non-dissipative for all wave numbers for all stable
Courant numbers.

\FloatBarrier

\begin{figure}[h]
\centering
          \subfigure[Courant number $\alpha = 0.25$]{\epsfig{figure=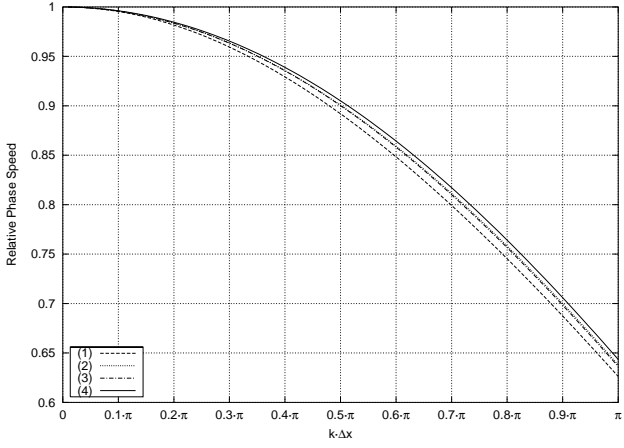,width=3.35in}\label{fig:11a}}
          \subfigure[Courant number $\alpha = 0.50$]{\epsfig{figure=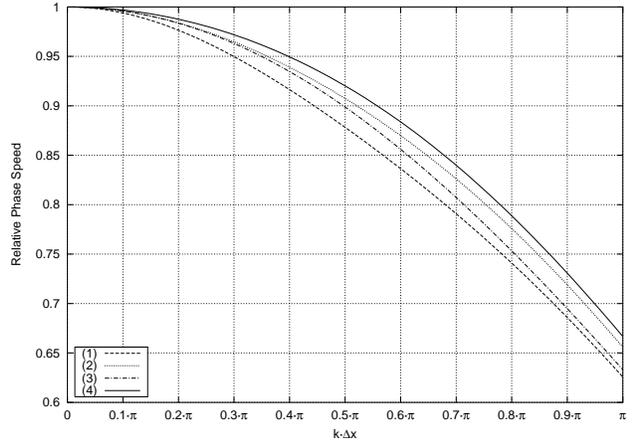,width=3.35in}\label{fig:11b}}
          \subfigure[Courant number $\alpha = 0.75$]{\epsfig{figure=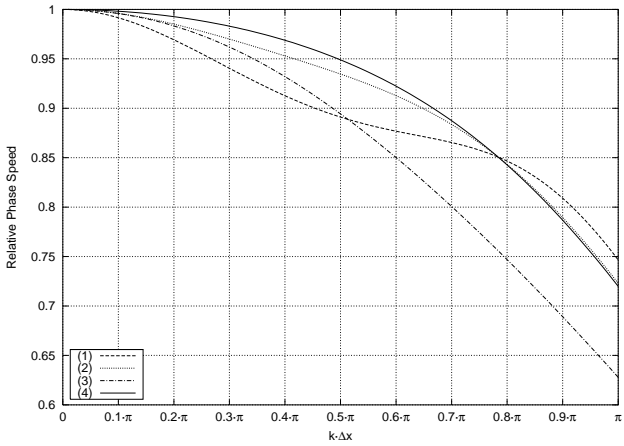,width=3.35in}\label{fig:11c}}
          \subfigure[Courant number $\alpha = 1.00$]{\epsfig{figure=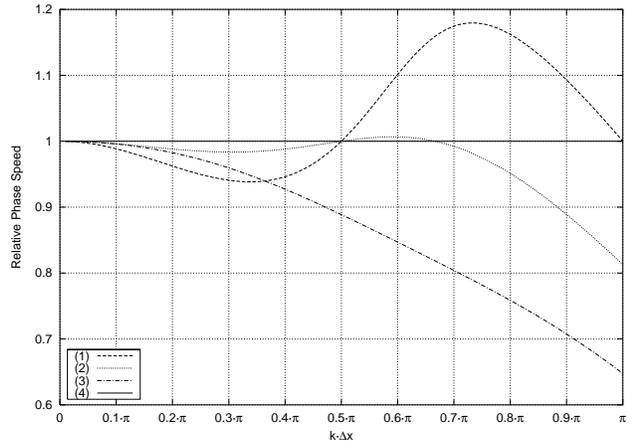,width=3.35in}\label{fig:11d}}
\caption{{\it Wave speeds for the four schemes for various Courant numbers relative to the analytic
          wave speed. Legend is (1) = ICN, (2) = RK3, (3) = RK4 and (4) = CFLN. 
\label{fig:11}}}
\end{figure}
Figures \ref{fig:11} compares the dispersion errors for the four schemes for various
Courant numbers as calculated from an Von Neumann analysis for equation \eqref{eq:linear}. For small Courant
numbers (fig. \ref{fig:11a}), it is seen that the schemes behaves 
quite similarly (in fact, for $\alpha \rightarrow 0$ the dispersion errors for the four schemes
converge to the same line), but a 
closer examination shows that the CFLN scheme has the smallest dispersion error,
then follows the RK3 and RK4 schemes respectively and finally the ICN scheme
which exhibits the largest dispersion errors. Also, dispersion errors are largest for large
wave numbers and going to zero in the limit $k \rightarrow 0$ for all schemes.\\
The dispersion errors for the ICN and RK4 schemes are increased monotonically for increasing Courant
numbers at smaller wave numbers, while the dispersion errors for these schemes show some non-monotonic
behavior at high wave numbers. As can be seen from figure \ref{fig:11d}, the ICN scheme in the limit
of $\alpha = 1.00$ exhibits positive dispersion errors for high wave numbers, i.e. the numerical
solution is propagating faster then the analytic solution\footnote{However, this is unlikely to be
  of numerical importance due the high damping for ICN in the high wave number / high Courant number
  limit.}. 

The dispersion errors for the RK3 and CFLN schemes conversely are minimized for their respective
maximal stable Courant numbers. The CFLN scheme in this limit has zero dispersion errors, while the
RK3 still has a non-vanishing (but minimized) dispersion error. 
\FloatBarrier

\section{\bf  Von Neumann stability analysis of the non-linear
  wave equation} 
\label{sec:nlwave}
We are interested in the behavior of the schemes in the non-linear regime.
In this section we present the results of a Von Neumann stability
analysis of the four schemes applied to a local linearizion of the non-linear wave equation,
equation \eqref{eq:nlwe2}, presented in section \ref{subsec:nlwave}.

\subsection{\bf   The spatial constant solutions}
\label{sec:2.01}
\setlength{\parindent}{0em} \setlength{\parskip}{1.5ex plus 0.5ex}
\begin{figure}[h]
\centering
          \subfigure[Decaying solution \eqref{eq:constant}.
          ]{\epsfig{figure=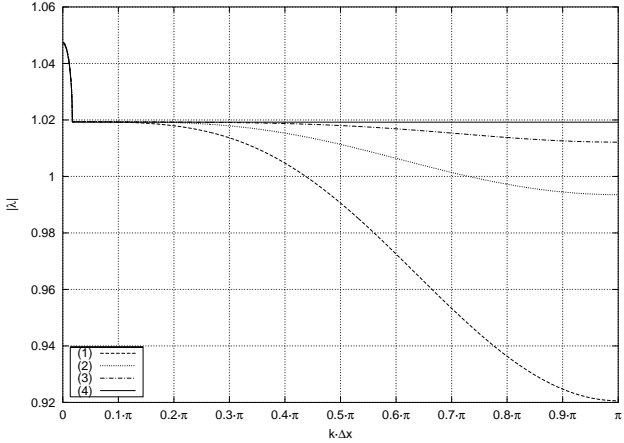,
          width=3.35in}\label{fig:sqrta}}    
          \subfigure[Growing solution \eqref{eq:constant}.
          ]{\epsfig{figure=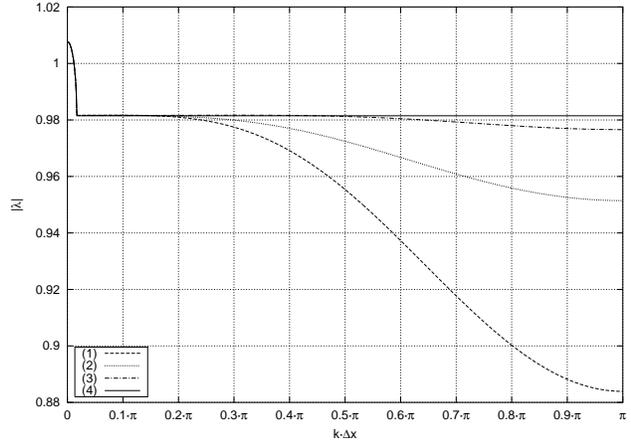, 
          width=3.35in}\label{fig:sqrtb}}
\caption{{\it Amplification factors for solution \eqref{eq:constant} as a function of $k\cdot\Delta x$. (a) is decaying ($C_2 = -1$)
          and (b) is growing ($C_2 = 1$). Courant number is $\alpha
    =0.5$ and $\Delta x=\frac{\pi}{16}$. Legend is (1) = ICN, (2) = RK3, (3) = RK4 and
    (4) = CFLN. 
\label{fig:sqrt}}}
\end{figure}
Figures \ref{fig:sqrt} shows a typical plot of amplification factors versus $k\cdot \Delta x$ for the
spatially constant solution
\eqref{eq:constant} of equation \eqref{eq:nlwe2} for decaying and growing solutions
respectively, for the four schemes. Solution parameters are $C_1=2$ for both plots and $C_2 = -1 $ and $C_2 = 1$ for
figure \ref{fig:sqrta} and \ref{fig:sqrtb} respectively. From the perturbation
analysis in subsection \ref{sec:linear_analysis} we expect to observe an amplification for
all wave numbers when solution \eqref{eq:constant} is decaying. For growing solutions we expect to
see a more modest amplification at
small wave numbers and damping at higher wave numbers.
Looking at figures \ref{fig:sqrt}, we observe that at small wave numbers, all schemes agree with
this prediction, i.e. we see a strong amplification at small wave numbers for the decaying solution
and a smaller amplification for the growing solution. At high wave numbers the CFLN scheme agrees
well with the perturbation analysis and we see an amplification at the Nyquist frequency, while the 
method of lines schemes all shows various degrees of damping, consistent with the results from
section \ref{sec:linear}. The amount of damping is dependent on the Courant number as in the
linear case, but also upon the spatial step size, $\Delta x$. By choosing a smaller $\Delta x$, for
a fixed Courant number, $\Delta t$ decreases proportionally, by which any amplification also becomes
smaller. However, with the amplification factor moving closer to $|\lambda |=1$, the damping for the
method of lines schemes are shifted proportionally, i.e. the method of lines schemes may go from
showing amplificative behavior to damping behavior at the Nyquist frequency, with the choice of a
smaller $\Delta x$. We see this effect in figure \ref{fig:sqrta} for the ICN scheme. If we for the
same solution had chosen a sufficiently small $\Delta x$, the other method of lines schemes would
also have shown damping behavior at the Nyquist frequency. The qualitative behavior of the CFLN
scheme, in contrast, is unaffected by the choice of $\Delta x$ for reasonably small $\Delta x$. For
very large $\Delta x$, all the schemes shows abnormal behavior which is caused by the fact that the
base solution in this case changes much more rapid then the temporal resolution allows for.

\subsection{\bf   The exponential solutions}
\label{sec:2.1}

\begin{figure}[h]
\centering
          \subfigure[Decaying solution \eqref{eq:exponential}]
          {\epsfig{figure=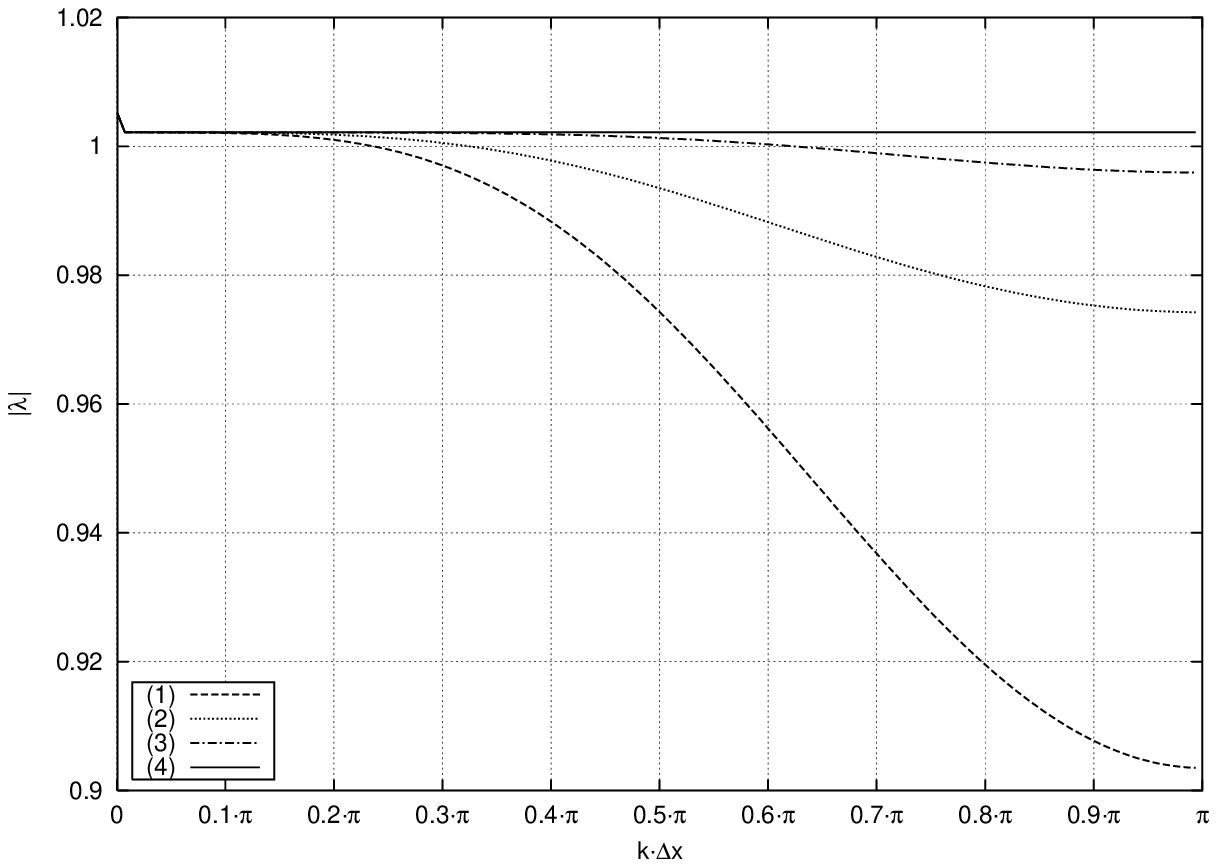, 
          width=3.35in}\label{fig:expa}}
          \subfigure[Growing solution \eqref{eq:exponential}]
          {\epsfig{figure=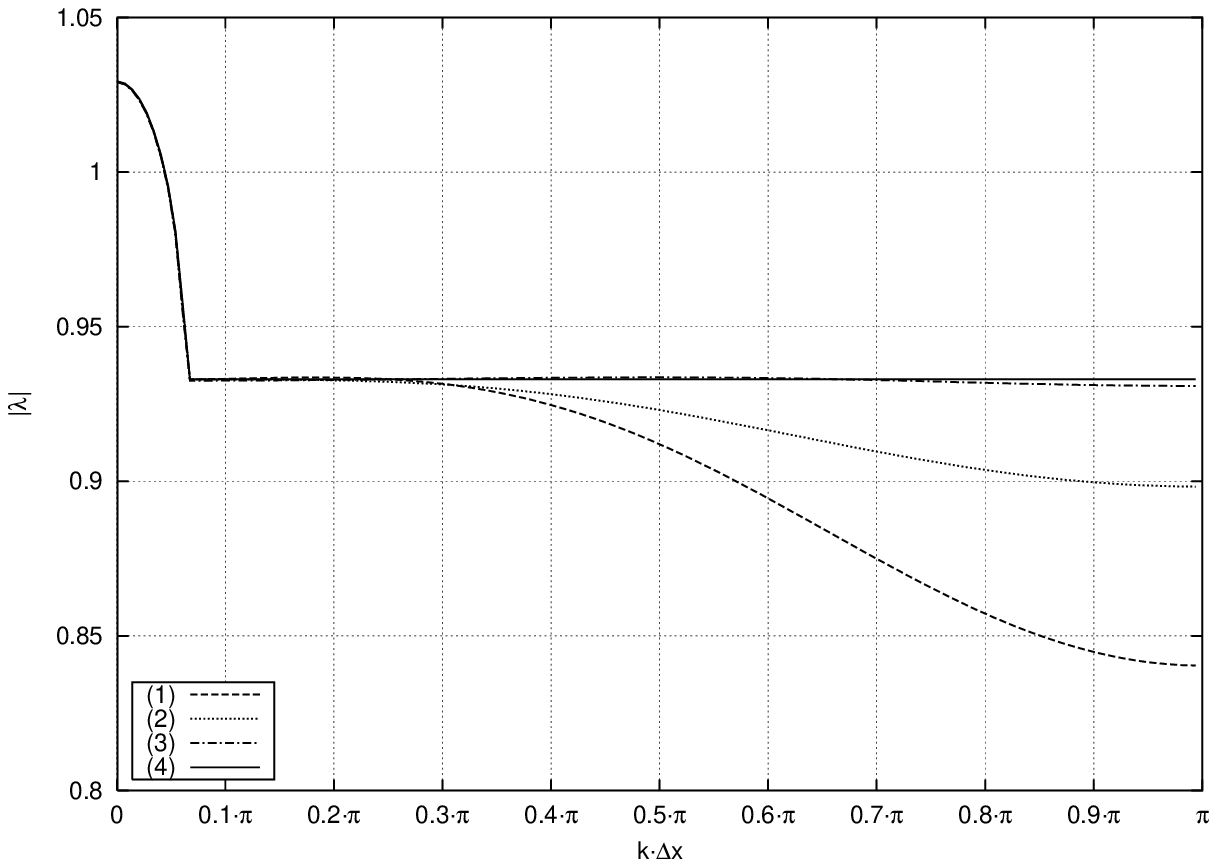,
          width=3.35in}\label{fig:expb}}     
\caption{{\it Amplification factors for solution \eqref{eq:exponential} as a function of
          $k\cdot\Delta x$.  (a) is decaying and (b) is growing. Courant number is $\alpha
    =0.5, \Delta x=\frac{\pi}{16}$ and $C=1$. Legend is (1) = ICN, (2) = RK3, (3) = RK4 and
    (4) = CFLN. 
\label{fig:exp}}}
\end{figure}

Figures \ref{fig:exp} shows a typical plot of the amplification factors for the four
schemes under investigation for the exponentially growing and decaying solution \eqref{eq:exponential}
to equation \eqref{eq:nlwe2} as a function of $k\cdot \Delta x$ with solution parameter
$C=1$. We see that the plots are 
very similar to figures \ref{fig:sqrt}. For the decaying solution (figure \ref{fig:expa}) we see
that CFLN is an amplificative scheme whereas the method of lines schemes displaying certain 
amount of dumping.
We note that in order to reproduce a 
correct behavior of perturbed analytic solutions the scheme must have the amplification 
number greater than one. However, in numerical simulations it may be better to dump growing large wave number (short 
wavelength) perturbations instead of trying to faitfully reproduce them. 

For small wave numbers both the decaying and growing solutions are indicating amplificative
behavior. We note that solution \eqref{eq:exponential} is spatially dependent, hence the spatially
constant perturbation analysis from subsection \ref{sec:linear_analysis} is not valid in this
regime. Nevertheless, comparing with figures \ref{fig:sqrt} where we see (and expect) the same
behavior it seems plausible that the amplificative behavior seen in the figures is an expected
behavior.

As a closing remark to this section, we note that all schemes and solutions has been tested in the
limit $\Delta t \rightarrow 0$ to investigate the stability properties of the schemes in the
non-linear regime and we have found that all schemes agree with the stability intervals for the
linear case (table \ref{tab:stableinterval}) for all solutions.

\FloatBarrier

\section{\bf \Large Numerical Tests}
\label{sec:4}
\begin{figure}[h]
\begin{center}
  \epsfig{figure=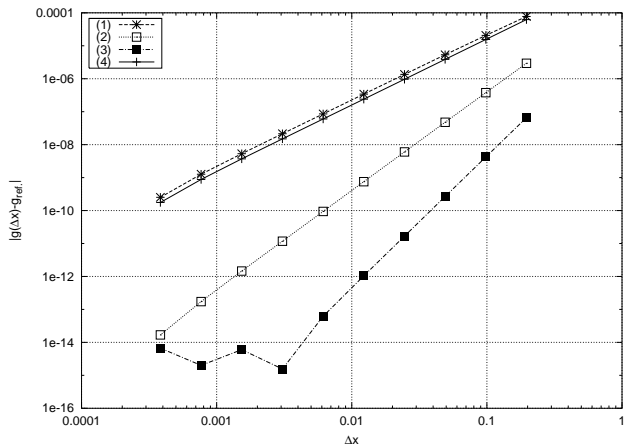,width=3.35in}
\caption[]{{\it  
    Absolute error between simulation at specified $\Delta x$
    vs. simulation at $\Delta x = \frac{\pi}{2^{14}}$ for Courant number $\alpha = 0.50$. Legend is
    (1) = ICN, (2) = RK3, (3) = RK4 and 
    (4) = CFLN. 
\label{fig:convergence_decsqrtt}}}
\end{center}
\end{figure}

We have done numerical testings to verify the analytic results presented in the preceding
sections. Figure \ref{fig:convergence_decsqrtt} shows the convergence of solution
\eqref{eq:constant} to equation \eqref{eq:nlwe2}. Plotted on the vertical axis is the absolute error
of the
central point in the domain between a simulation with the spatial step specified on the horizontal
axis and a reference simulation. The reference simulation is a high resolution simulation with a
resolution twice that of the leftmost point in the 
plot. The constants in solution \eqref{eq:constant} were set to $C_1 = 2$ and $C_2=-1$, the
simulations were run for a time $t=\pi /4$ in a domain of size $2\pi$ and the Courant number of the
simulations was fixed at $\alpha = 0.50$.

\FloatBarrier

From the figure we see that the CFLN and ICN schemes are showing second order convergence, the RK3
scheme shows third order convergence. The RK4 scheme shows fourth order convergence at large $\Delta
x$, but flattens out a around $10^{-15}$ and for small step sizes. The flattening is caused by
machine precision errors affecting the solutions. 
The convergence rates are what should be expected for spatially constant solutions. This solution does not
introduce any truncation errors associated with spatial discretization. The figure thus shows
convergence rates in agreement with the 
truncation error due to the temporal discretization of the schemes (second order CFLN and ICN, third
order RK3, and fourth order RK4).
Identical convergence tests for the exponential solutions \eqref{eq:exponential} shows second order
convergence for all schemes due to the second order spatial finite difference operator we have used
in the schemes to calculate spatial derivatives.

\begin{figure}[h]
\begin{center}
  \epsfig{figure=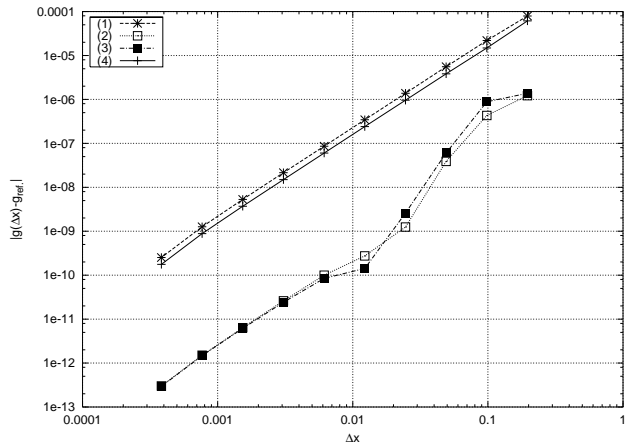,width=3.35in}
\caption[]{{\it  
    Absolute error between perturbed simulation at specified $\Delta x$
    vs. simulation at $\Delta x = \frac{\pi}{2^{14}}$ for Courant number $\alpha = 0.50$. Perturbation
    is a moving sinusoidal wave with wave number $k=16$ and initial amplitude $A_0 = 10^{-6}$. 
    Legend is (1) = ICN, (2) = RK3, (3) = RK4 and (4) = CFLN. 
\label{fig:convergence_perturbed}}}
\end{center}
\end{figure}

To test analytic predictions of the behavior of a perturbed solution we have made numerical tests
of base solutions perturbed by small amplitude sinusoidal perturbations,
$g(x,t)=g_0(x,t)+\tilde{g}(x,t,k)$, with wave number $k$.
Figure \ref{fig:convergence_perturbed} is identical to figure \ref{fig:convergence_decsqrtt}, except
that now base solution \eqref{eq:constant} is perturbed by a moving sinusoidal wave with
wave number of $k=16$ (corresponding to the Nyquist frequency for the rightmost point) and initial
amplitude $A_0 = 10^{-6}$, all other parameters are as for figure \ref{fig:convergence_decsqrtt}.

\FloatBarrier

We see from the figure that the perturbed solution is converging for all schemes. For higher
resolutions the convergence is at least second order. Similar convergence rates are seen for
analogous simulations with other base solutions.

\begin{figure}[h]
\begin{center}
  \epsfig{figure=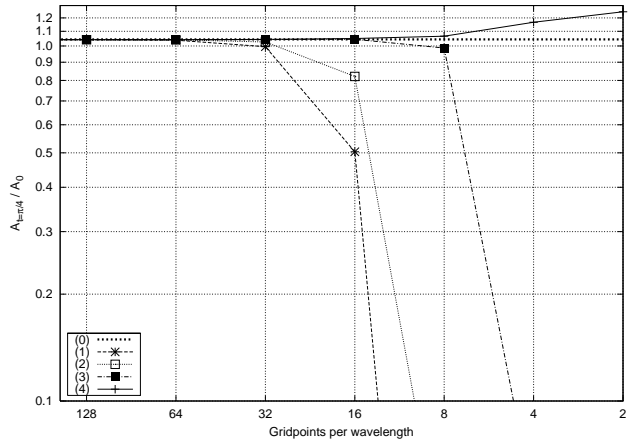,width=3.35in}
\caption[]{{\it  $A_t / A_0$ of sinusoidal perturbation at $t=\pi /4$ for base solution
    \eqref{eq:constant} versus various wave numbers for a fixed $\Delta x =
 \frac{\pi}{2^{13}}$ and $\alpha =0.50$. Legend is (0) = Analytic prediction, (1) = ICN, (2) = RK3, (3) = RK4 and (4) = CFLN. 
\label{fig:amplification_convergence}}}
\end{center}
\end{figure}

Figure \ref{fig:amplification_convergence} shows the ratio of the final amplitude ($A_t$) of a
perturbation at time $t=\pi /4$ to its initial amplitude ($A_0 = 10^{-6}$) as a function of the
resolution of the perturbed wave for simulations with base solution \ref{eq:constant} with $C_1=10$
and $C_2=-1$. Courant $\alpha = 0.50$ and  $\Delta x=\frac{\pi}{2^{13}}$ was used in all simulations.
The rightmost data point corresponds to a perturbation being at the Nyquist frequency. The
analytical prediction is that 
\[A_{t=\pi /4}/A(0) = \exp{-\frac{C_2\cdot t}{2C_1-2C_2\cdot t}}\approx 1.0435\]
according to
\eqref{eq:perturbation_analysis_amp}. This prediction is also shown on the figure as line (0).

From this figure we see that at high resolutions all schemes produce the same growth of amplitude of
perturbations, which is in agreement with the analytical prediction. As the resolution decreases the
results for all schemes for all schemes begin to deviate from the analytical value. Among the method
of lines schemes, the RK4 is the most accurate and is able to
reproduce the correct behavior of the perturbation up to the resolution of 16 grid points per
wavelength. CFLN also requires only 16 grid points per
wavelength to reproduce the correct behavior. The other two schemes requires from 32 to 64
grid points per wavelength. The major difference between the CFLN and the method of lines schemes is that at
low resolutions, CFLN is amplifying the perturbations more than it should according to the analytical
predictions, whereas the method of lines schemes are damping. The RK4 is the least damping and the
most accurate of these schemes.
The behavior of the schemes at low resolutions is fully consistent with the results of the Von
Neumann stability analysis, presented in section \ref{sec:2.01}. According to that analysis the CFLN
scheme must 
be amplifying at all wavelengths whereas the method of lines schemes must be damping.

\section{\bf \Large Discussion and conclusion}
In this paper we studied the properties of four numerical schemes, CFLN, ICN, RK3 and RK4, applied to
a non-linear scalar wave equation \eqref{eq:nlwe}. This equation has a number of non-trivial
analytic solutions whose properties, including stability, were studied and summarized in
section \ref{sec:theory}. We carried out the Von Neumann stability analysis of the schemes and studied
their phase and amplitude errors.  Finally we carried out numerical experiments and compared the
results of those experiments with the perturbation analysis of the equation and the Von Neumann
stability analysis of the schemes. 

We find that all four schemes presented in this paper are stable and converge with the second order
accuracy. The stability range for the schemes were determined and are presented in table
\ref{tab:stableinterval}. Those ranges are valid for both linear and non-linear solutions. 

For non-linear
schemes we checked that the amplification factor $|\lambda|$ is less than $1+O(\Delta t)$ for sufficiently
small $\Delta t$.  

With respect to the dissipation errors, we find that the CFLN has the least amount of dissipation. The
ICN scheme has the largest dissipation errors. The RK3 and RK4 are intermediate. For the method of
lines schemes we find that damping is behaving non-monotonically with the increase of the Courant
number.

With respect to phase errors, the schemes can be arranged in the sequence CFLN RK3, RK4 and ICN,
with the CFLN having the least amount of errors and ICN having the largest. 

We find that CFLN scheme requires the least amount of operations per time step, whereas the RK4
requires the largest amount. 

Ideally a numerical scheme is preferable which has the minimal phase and amplitude errors, and is
computationally inexpensive. It is also of practical importance to have a scheme which will damp the
high frequency perturbations at and close to the Nyquist frequency. Otherwise truncation errors at
the highest frequency will remain within the computational domain, may be amplified and may
eventually spoil the solution. None of the schemes discussed in this paper satisfy all those
criteria. We think that the RK4 scheme should be preferred to other schemes because of its damping
properties at the Nyquist frequency and the minimal amount of errors consistent with this
property. On the other hand, the amount of dissipation in the RK4 scheme is dependent upon Courant
number and is not controllable. If not for the damping properties, the CFLN scheme is the most
accurate and cost effective. It would be of great interest to develop a version of this scheme with
a controllable filter, for damping the minimum amount of perturbations at a narrow range of
frequencies near the Nyquist frequency.

\bigskip
\noindent
This work was supported in part by Danmarks
Grundforskningsfond through its support for establishment of the Theoretical Astrophysics Center and
by the Danish SNF Grant 21-03-0336. We
thank A. Doroshkevich and R. Takahashi for useful discussions. The authors thank Caltech for hospitality
during their visits.

\FloatBarrier

\end{document}